\pdfoutput=1
\documentclass[twocolumn,superscriptaddress,aps,preprintnumbers,amsmath,amssymb,prd,nofootinbib]{revtex4-2}

\usepackage{amsmath}
\usepackage{amssymb}
\usepackage{amsfonts}
\usepackage{graphicx}
\usepackage{xcolor}
\usepackage{xfrac}
\usepackage{comment}
\usepackage{pifont}
\usepackage{physics}
\usepackage{fourier}
\usepackage{hyperref}
\usepackage{bm}
\usepackage{enumitem}

\definecolor{rossoferrari}{HTML}{D9073D}
\definecolor{mediumblue}{HTML}{0000CD}
\definecolor{forestgreen}{HTML}{228B22}
\definecolor{desy_blue}{HTML}{009EE2}
\definecolor{desy_orange}{HTML}{FD8800}
\definecolor{light_pink}{rgb}{1,0.4,0.4}
\definecolor{light_blue}{rgb}{0.284602,0.317763,0.963947}
\hypersetup{setpagesize=false,bookmarksnumbered=true,bookmarksopen=true, colorlinks=true,linkcolor=light_blue,urlcolor=rossoferrari,citecolor=rossoferrari,linktocpage=false}

\usepackage{slashed}

\renewcommand{\thefootnote}{\fnsymbol{footnote}}

\newcommand{\bea}{\begin{array}}
\newcommand{\eea}{\end{array}}
\newcommand{\beq}{\begin{eqnarray}}
\newcommand{\eeq}{\end{eqnarray}}

\newcommand{\lmk}{\left(}  
\newcommand{\rmk}{\right)}
\newcommand{\lkk}{\left[}  
\newcommand{\rkk}{\right]}

\newcommand{\del}{\partial}

\def\eq#1{Eq.~(\ref{#1})}

\definecolor{orange}{RGB}{255,100,0}
\definecolor{rosepink}{RGB}{248,100,100}

\begin{document}

\title{
A new regularization scheme for the wave function of the Universe in the Lorentzian path integral
}

\author{Masaki Yamada}
\email{m.yamada@tohoku.ac.jp}
\affiliation{Department of Physics, Tohoku University, Sendai, Miyagi 980-8578, Japan}

\preprint{TU-1286}

\date{\today}

%%%%%%%%%%%%%%%%%%%%%%%%%%%%%%%%%%%%%%%%%%%%%%%%%%%%%%%%%%%%%%%%%%%%%%%%%%%%%%%%%%%%%%%%%%%%%%%%%%%%

\begin{abstract}
\noindent
The Lorentzian path integral for the wave function of the Universe is only conditionally convergent and thus requires a well-defined prescription. The Picard-Lefschetz approach ensures convergence through contour deformation, but it has been argued that this leads to unsuppressed perturbations due to relevant saddle points residing in the region ${\rm Im}N>0$. As an alternative, we propose a simple regulator for the lapse integral in minisuperspace. Specifically, we impose the vanishing initial size of the Universe via a delta function, represented as a narrow Gaussian of width $\sigma$, and take the limit $\sigma \to 0$ only after performing the functional integrations. This regulator has a clear physical interpretation: it corresponds to a vanishingly small quantum uncertainty in the initial size of the Universe. For any fixed $\sigma > 0$, the lapse integral is absolutely convergent along (or slightly below) the real axis, and no excursion into the region ${\rm Im}N>0$ is required. We further argue that the initial wave function for scalar and tensor perturbations should be incorporated in the Lorentzian path integral formalism, and we show that these perturbations are then appropriately suppressed. A purely Lorentzian path integral thus yields the tunneling wave function with suppressed perturbations. We also demonstrate that the Hartle-Hawking wave function can be reproduced by choosing a contour for the lapse integral extending from $-\infty$ to $+\infty$ that passes below the singularity near the origin.
\end{abstract}

\maketitle

\renewcommand{\thefootnote}{\arabic{footnote}}
\setcounter{footnote}{0}

%%%%%%%%%%%%%%%%%%%%%%%%%%%%%%%%%%%%%%%%%%%%%%%%%%%
\section{Introduction}
%%%%%%%%%%%%%%%%%%%%%%%%%%%%%%%%%%%%%%%%%%%%%%%%%%%

The origin of the universe remains a central problem in theoretical cosmology. 
If the initial size of the universe is vanishingly small, quantum-gravitational effects should govern the birth of spacetime. In particular, the spacetime metric itself must be treated as a quantum variable~\cite{Vilenkin:1983xq,Vilenkin:1984wp,Hartle:1983ai,Linde:1983mx,Rubakov:1984bh,Vilenkin:1984wp,Zeldovich:1984vk}. 
Within quantum cosmology, two well-known proposals aim to define the ``wave function of the universe'': the tunneling proposal~\cite{Vilenkin:1984wp,Vilenkin:1984wp} and the Hartle--Hawking (HH) no-boundary proposal~\cite{Hartle:1983ai}. 
These proposals may be formulated either via boundary conditions on the Wheeler--DeWitt (WdW) equation~\cite{DeWitt:1967yk,Vilenkin:1986cy,Vilenkin:1987kf} or via different contour choices in gravitational path integrals~\cite{Hartle:1983ai,Vilenkin:1984wp,Halliwell:1988ik,Halliwell:1989dy,Brown:1990iv,Vilenkin:1994rn,Feldbrugge:2017kzv}.

Originally, the HH no-boundary proposal was formulated within the Euclidean path integral approach. 
However, this suffers from the so-called conformal problem, which states that the Euclidean action is unbounded from below, particularly under conformal deformations of the metric~\cite{Gibbons:1978ac}. 
At best, one may hope that the Euclidean path integral can be approximated semiclassically by saddle points (i.e., classical solutions), assuming that the conformal problem is resolved by yet-unknown physics beyond the semiclassical regime.

The conformal problem can seemingly be avoided in the Lorentzian path integral formulation, since the integrand remains oscillatory even beyond the semiclassical level~\cite{Halliwell:1988ik,Feldbrugge:2017kzv}. 
Although the Lorentzian integral is only conditionally convergent, recent progress employing Picard--Lefschetz theory has made it possible to deform the integration contour onto steepest-descent thimbles, rendering the integral absolutely convergent and selecting the relevant semiclassical saddle points unambiguously~\cite{Feldbrugge:2017kzv} (see also Refs.~\cite{Honda:2024aro,Chou:2024sgk}). 
This approach yields a semiclassical exponent consistent with the tunneling wave function and leads to unsuppressed (inverse-Gaussian) fluctuations for linear perturbations~\cite{Feldbrugge:2017fcc} (see also Ref.~\cite{Halliwell:1989dy} for an earlier study). 
Attempts to reproduce the HH wave function by introducing intrinsically complex lapse contours~\cite{DiazDorronsoro:2017hti,DiazDorronsoro:2018wro,Janssen:2019sex} inevitably lead to additional nonperturbative contributions, which again result in unsuppressed fluctuations~\cite{Feldbrugge:2017mbc,Feldbrugge:2018gin}. 
Alternative approaches impose suitable (Robin-type) boundary conditions on the perturbations~\cite{Vilenkin:2018dch,Vilenkin:2018oja,Wang:2019spw} or on the scale factor~\cite{DiTucci:2019dji,DiTucci:2019bui,Ailiga:2023wzl,Ailiga:2024mmt,Ailiga:2024wdx}, aiming to recover either the tunneling or HH wave functions while maintaining well-behaved perturbations. 
Other proposals include the consideration of off-shell instantons in loop quantum cosmology~\cite{ Bojowald:2018gdt,Bojowald:2020kob}, 
analytic continuation from negative to positive potentials~\cite{Lehners:2021jmv}, 
and modifications of the dispersion relation motivated by trans-Planckian physics~\cite{Matsui:2022lfj}. 
An alternative definition of the wave function based on saddle points satisfying regularity conditions has also been discussed in Ref.~\cite{Halliwell:2018ejl}.

In this paper, we propose a new prescription to define the Lorentzian (or nearly Lorentzian) path integral by introducing a regularization of the conditionally convergent lapse integral. 
Our idea is to insert a unity that enforces a vanishing initial size of the universe, expressed in terms of $q \equiv a^2$ with $a$ being the scale factor:
\begin{equation}
 1 \;=\; \int\! {\rm d} q_0\, \delta(q_0)\,,
\end{equation}
and to represent the delta function by a narrow Gaussian of width $\sigma$:
\begin{equation}
 \delta(q_0) \;=\; \lim_{\sigma\to 0}\, \frac{1}{\sqrt{2\pi}\,\sigma}\,
 \exp\!\left[-\,\frac{q_0^2}{2\sigma^2}\right] \,.
\end{equation}
We then exchange the order of operations: first perform the path integrals at fixed $q_0$, then carry out the Gaussian-weighted $q_0$ integration, and finally take the limit $\sigma \to 0$. 
This defines the propagator from $q = q_0 = 0$ to $q = q_1$ as
\begin{equation}
G(q_0{=}0\,;q_1) \;=\; 
\lim_{\sigma\to 0}\,
\int \frac{{\rm d}q_0}{\sqrt{2\pi}\,\sigma}
\int_{q_0}^{q_1}\!\! \mathcal{D}g\,\mathcal{D}\phi\;
\exp\!\Big[iS[g,\phi]-\frac{q_0^2}{2\sigma^2}\Big] \,,
\label{G}
\end{equation}
where $S$ is the Einstein--Hilbert action plus matter. 
In a minisuperspace, $g$ collectively denotes $q$ and the lapse $N$. 
We show that, for any fixed, finite $\sigma$, the integral over the lapse $N$ converges along the \emph{real} axis without requiring an ad hoc contour deformation. 
In this sense, the exponential factor acts as a regulator, with $\sigma$ serving as a cutoff that is removed in the limit $\sigma \to 0$. 
When the original lapse domain is taken as $N \in (0,\infty)$, the resulting amplitude reproduces the tunneling wave function. 
We numerically verify that our regularization scheme is consistent with the saddle-point approximation in the $\sigma \to 0$ limit.

There are several advantages to employing a purely Lorentzian path integral without deforming the contour in the complex $N$ plane.
First, the fluctuations of the scale factor and scalar/tensor perturbations remain convergent, and their path integrals can be evaluated beyond the semiclassical approximation. This improvement stems from the fact that the formulation is free from the conformal problem that arises in the Euclidean path integral.
Second, the initial wave function of the scalar/tensor perturbations must be included, which ensures that these modes are appropriately suppressed on the final time slice.
Because we consider a purely Lorentzian path integral, the propagator itself is generally distinct from the wave function. 
In particular, when scalar or tensor perturbations are included, the wave function at a final time slice is obtained by convoluting the propagator with an initial wave function. 
We argue that the initial wave function should be chosen to be consistent with the Bunch--Davies vacuum, since in the limit of vanishing scale factor the mode equation reduces to that of a massless conformal scalar field, which admits a positive-frequency mode function independent of the time dependence of the scale factor. 
Consequently, the perturbations are Gaussian and appropriately suppressed.

Beyond the purely Lorentzian path integral, the HH wave function can also be recovered by appropriately deforming the integration contour for the lapse function. 
Although the conformal problem is absent only for the strictly Lorentzian integral, our regularization scheme for the lapse integral remains applicable to any contour asymptotically approaching $N=\pm\infty$, which we refer to as a nearly Lorentzian path integral. 
If the original lapse contour runs from $-\infty$ to $+\infty$ and passes below the singularity near the origin, we find that the resulting path integral yields the HH wave function.%
\footnote{
\label{footnote1}
Since the contour passes through $\operatorname{Im}N<0$, the integral is not strictly along the real axis. Nevertheless, it remains effectively Lorentzian because the deformation is confined to a small neighborhood of $N=0$, which can be made arbitrarily small as $\sigma \to 0$.}
In contrast to Picard--Lefschetz-based prescriptions, where relevant thimbles typically lie above the real axis, our integral remains convergent even when restricted to contours along or below the real axis. 
As a result, the scalar and tensor perturbation integrals remain well behaved.

The organization of this paper is as follows. 
In Sec.~\ref{sec:regulator}, we present our regularization method and evaluate the path integral for the scale factor and scalar perturbations. 
In Sec.~\ref{sec:lapse}, we examine the convergence behavior of the lapse integral in the complex plane and demonstrate that it converges along the real axis. 
We then compute the lapse integral analytically and numerically, taking the $\sigma \to 0$ limit. 
In Sec.~\ref{sec:HH}, we consider an alternative contour that reproduces the HH wave function, paying particular attention to the singular curve in the complex plane. 
We summarize and discuss our results in Sec.~\ref{sec:discussions}. 
Finally, the Appendix clarifies how to obtain the wave function of scalar and tensor perturbations from the propagator and the initial wave function within the Lorentzian path-integral formalism.

%%%%%%%%%%%%%%%%%%%%%%%%%%%%%%%%%%%%%%%%%%%%%%%%%%%
\section{Lorentzian path integral with a regulator}
\label{sec:regulator}
%%%%%%%%%%%%%%%%%%%%%%%%%%%%%%%%%%%%%%%%%%%%%%%%%%%

We consider the Lorentzian path integral for gravity coupled to a scalar or tensor field (collectively denoted by $\phi$) in a minisuperspace model. 
The metric is taken as~\cite{Halliwell:1988ik}
\begin{equation}
{\rm d}s^2 \;=\; -\,\frac{N^2}{q(t)}\,{\rm d}t^2 \;+\; q(t)\, {\rm d}\Omega_3^2 \,,
\end{equation}
where $q(t)\equiv a^2(t)$ is the squared scale factor and $N(t)$ is the lapse function. 
Fixing the constant-lapse gauge $\dot N=0$, the gravitational action reduces to
\begin{align}
S_g[q,N]
&= \int {\rm d}^4x\,\sqrt{-g}\,\lmk \frac{R}{2}-\Lambda \rmk
\\
&= 6\pi^2 \int_0^1 {\rm d}t \,\Big[ -\,\frac{\dot q^{\,2}}{4N} + N\,(1 - H^2 q) \Big] \,,
\label{S0q}
\end{align}
with $\Lambda=3H^2$, where we set $8\pi G =1$. 
The propagator from $(q_0,\phi_0)$ at $t = 0$ to $(q_1,\phi_1)$ at $t = 1$ is given by 
\begin{equation}
G(q_1,\phi_1; q_0,\phi_0)
\;=\; \int {\rm d}N \int_{q_0}^{q_1} \!\! \mathcal{D}q \; e^{\,i S_g[q,N]}\;
\int_{\phi_0}^{\phi_1} \!\! \mathcal{D}\phi \; e^{\,i S_\phi[\phi;q,N]}\,, 
\end{equation}
where $S_\phi$ denotes the action for scalar/tensor perturbations, specified below. 

As explained in the Introduction, rather than imposing $q_0\equiv q(0)=0$ strictly, we insert a unity and exchange the order of integrations and limits to define the propagator:
\begin{equation}
G(q_1, \phi_1; q_0,\phi_0) \;=\;
\lim_{\sigma\to 0}\int {\rm d}N
\int \frac{{\rm d}q_0}{\sqrt{2\pi}\,\sigma}
\int_{q_0}^{q_1} \!\!\mathcal{D}q
\int_{\phi_0}^{\phi_1}\!\!\mathcal{D}\phi\;
e^{\,i\,(S_g'+S_\phi)}\,, 
\end{equation}
with the regulated gravitational action
\begin{equation}
S_g'[q,N;q_0] \;\equiv\; S_g[q,N] \;+\; i\,\frac{q_0^{\,2}}{2\sigma^2}\,.
\end{equation}
As we will see in Sec.~\ref{sec:lapse}, 
for any fixed, finite $\sigma$, the Gaussian factor renders the lapse integral convergent \emph{along the real axis}, so that no \emph{ad hoc} contour deformation is required to define the propagator. 
Depending on the original integration domain for $N$, this construction reproduces either the tunneling or the HH wave function, as shown below.

We note that the path integral over $\phi$ should be performed prior to that over $N$, since the action $S_\phi$ also depends on $N$.
One might hope to evaluate it after integrating over $N$, assuming that the backreaction is negligible. However, care must be taken to ensure that the path integral over $\phi$ remains convergent for a general value of $N$ in the complex $N$ plane, and to recognize that the propagator itself cannot be directly identified as a wave function in the Lorentzian formulation.
To clarify these points, we evaluate the case of a massless conformally coupled field $\phi$, which serves as an analytically tractable example.

%-------------------------------------------------
\subsection{Path integral for the scale factor}
%-------------------------------------------------

We first neglect scalar/tensor perturbations and focus on the path integral for $q$. 
Since $S_g'$ is at most quadratic in $q$, the path integral is dominated by the classical trajectory obtained from the variational principle. 
Because $q_0$ is also integrated over, its value is determined dynamically. 
Varying $S_g'$ with fixed $q_1$ but arbitrary $\delta q_0$ yields
\begin{equation}
\delta S_g'
= 6\pi^2 \int_0^1 {\rm d}t\;\Big( \frac{\ddot q}{2N} - N H^2 \Big)\delta q
\;+\;\Big( \frac{3\pi^2}{N}\,\dot q_0 \;+\; i\,\frac{q_0}{\sigma^2}\Big)\delta q_0 \,.
\end{equation}
Hence, for the classical trajectory $q = q_{\rm c}(t)$ with $q_{\rm c}(0) = q_0$, the bulk equation of motion and the initial boundary condition become
\begin{equation}
\ddot q_{\rm c}(t) \;=\; 2 H^2 N^2,
\qquad
\dot q_{\rm c}(0) \;=\; -\, i\,\frac{N}{3\pi^2\sigma^2}\,q_{\rm c}(0) \,.
\label{eq:Robin}
\end{equation}
This boundary condition is of the Robin type, analogous to that introduced in Refs.~\cite{DiTucci:2019dji,DiTucci:2019bui,Ailiga:2023wzl,Ailiga:2024mmt,Ailiga:2024wdx}. 
Imposing $q_{\rm c}(1)=q_1$, the solution is
\begin{equation}
 q_{\rm c} (t) = H^2 N^2 t^2 + \frac{q_1 - H^2 N^2}{N + i 3 \pi^2 \sigma^2} N t + i 3\pi^2 \frac{q_1 - H^2 N^2}{N + i 3 \pi^2 \sigma^2} \sigma^2 \,. 
 \label{qt}
\end{equation}
Substituting \eqref{qt} into $S_g'$, the on-shell action becomes 
\begin{align}
S_g' [q_c, N; q_c(0)] &= \frac{6\pi^2}{N + i 3 \pi^2 \sigma^2} \left[
\frac{H^4}{12} N^4 + \left(1-\frac{1}{2}H^2 q_1\right) N^2 
\right.\nonumber\\
& \left.
-\frac{q_1^2}{4} + 3 \pi^2 i \sigma^2 N \lmk 1 - H^2 q_1 + \frac{1}{3} H^4 N^2 \rmk
\right] \,.
\nonumber\\
\label{SgN}
\end{align}

Alternatively, it is instructive to evaluate the path integral in a more direct manner. 
First, we compute the classical contribution of the path integral 
$\int_{q_0}^{q_1} \mathcal{D} q e^{i S_g[q,N]}$. 
The classical solution $\tilde{q}_c$ with boundary conditions $\tilde{q}_c(0) = q_0$ and $\tilde{q}_c(1) = q_1$ is given by 
\begin{equation}
 \tilde{q}_c(t) = H^2 N^2 t^2 + \lmk q_1 - q_0 - H^2 N^2 \rmk t + q_0\,.
 \label{eq:tildeqt}
\end{equation}
Substituting this into the action, the corresponding classical action reads
\begin{align}
 &S_g[\tilde{q}_c,N] \nonumber\\
 &= - \frac{\pi^2}{2 N} 
 \lkk 3 \lmk q_1 - q_0 \rmk^2 +6 N^2 \lmk H^2 (q_1+q_0) - 2 \rmk + H^4 N^2 \rkk \,.
\end{align}
We then evaluate the integral over $q_0$ with the Gaussian weight $e^{-q_0^2 / (2 \sigma^2)}$. 
Since the exponent is quadratic in $q_0$, the integral can be performed exactly:
\begin{equation}
 \int \frac{{\rm d}q_0}{\sqrt{2\pi}\,\sigma} e^{i S_g'[\tilde{q}_c,N; q_0]} = 
e^{i S_g'} \sqrt{\frac{N}{N + i 3 \pi^2 \sigma^2}} \,.
\end{equation}
The resulting classical action $S_g'$ coincides with \eqref{SgN}. 
Moreover, substituting the center of the Gaussian,
$q_0 = i 3 \pi^2 \sigma^2 (q_1 - H^2 N^2)/(N + i 3 \pi^2 \sigma^2)$,
into \eqref{eq:tildeqt} reproduces \eqref{qt}, which satisfies the boundary condition~\eqref{eq:Robin}. 
We emphasize that the Robin boundary condition is not imposed from the beginning. Rather, the classical trajectory that dominates the path integral is automatically the one that satisfies this Robin boundary condition.

The path integral over $q$ can now be evaluated by expanding around the classical trajectory:
\begin{equation}
 q = \tilde{q}_{\rm c}(t) + Q(t) \,,
\end{equation}
where $Q(t)$ represents Gaussian fluctuations satisfying $Q(0) = Q(1) = 0$. 
Integrating over all paths of $Q(t)$, we obtain
\begin{equation}
 \int_{Q(0)=0}^{Q(1) = 0} \mathcal{D} Q 
 \exp \lkk i c \int_0^1 {\rm d}t \dot{Q}^2 \rkk 
 = \sqrt{\frac{c}{\pi i }} \,,
 \label{eq:Gaussian}
\end{equation}
for $\Im c \ge 0$, 
with $c = - 6 \pi^2 /(4N)$. 
We note that the path integral includes Gaussian integrals that are well defined only for $\Im N \ge 0$.%
\footnote{
For the case of $\Im N = 0$, the path integral includes a Fresnel-type integral. 
Although it is only conditionally convergent, it can be defined by a suitable prescription, (for instance, $\lim_{R \to \infty} \int_{-R}^R e^{i a x^2} {\rm d}x = \sqrt{\pi i/a}$), which leads to Eq.~\eqref{eq:Gaussian}. 
}

We thus obtain 
\begin{equation}
\int \frac{{\rm d}q_0}{\sqrt{2\pi}\,\sigma}
\int_{q_0}^{q_1} \!\!\mathcal{D}q \;
e^{\,i\,S_g'} = \sqrt{\frac{3 \pi i }{2 (N + i 3 \pi^2 \sigma^2) }} e^{\,i\,S_g'[q_c, N; q_c(0)]} \,.
\end{equation}

%-------------------------------------------------
\subsection{Path integral for the scalar perturbations}
%-------------------------------------------------

Next, we consider the path integral for scalar and tensor perturbations, collectively denoted as $\phi$. 
We expand the field $\phi$ as 
\begin{align}
 \phi(x,t) = \sum_{n} \phi_n (t) Q_n(x) \,,
 \\
 \int Q_n(x) Q_{n'} {\rm d} \Omega_3 = \delta_{n n'} \,,
\end{align}
where $Q_{nlm}(x)$ are spherical harmonics on $S^3$ with integer $n$. 
Here and in what follows we omit the indices $l$ and $m$ for brevity. 
The tensor perturbations have $n \ge 3$, while the scalar perturbations have $n \ge 1$. 
We collectively denote both as $\phi$ and consider the general case, including possible nonminimal coupling to gravity. 
Later we will specialize to the case of a massless, conformally coupled scalar field.

The path integral can then be written as 
\begin{equation}
\int_{\phi_0}^{\phi_1}\!\!\mathcal{D}\phi\;
e^{\,i\,S_\phi} = 
\prod_n \int {\cal D}\phi_n e^{iS_n [\phi_n]} \,,
\end{equation}
with 
\begin{equation}
 S_n [ \phi_n] = \pi^2 \int_0^1 d t 
  \lkk \frac{q^2}{N} {\dot\phi}_n^2 - N \lmk (n^2-1) + q ( m^2 + \xi R) \rmk \phi_n^2 \rkk \,,
  \label{S_n}
\end{equation}
where we include a nonminimal coupling term proportional to $\xi R$. 
We also implicitly incorporate the Gibbons--Hawking boundary term and the term proportional to $\xi$ containing the extrinsic curvature on the boundary.

In general, the path integrals for the scalar field and the scale factor must be evaluated simultaneously because they satisfy coupled equations of motion. 
However, such a computation is typically intractable analytically and even numerically, except for special cases such as a conformally coupled massless scalar field. 
We therefore consider this case as an illustrative example of how to evaluate the path integral explicitly.

For a conformally coupled massless scalar field, the action can be rewritten as 
\begin{equation}
 S_n[\chi_n] = \pi^2 \int {\rm d} \eta 
 \lkk \frac{1}{N} \chi_n'^2 - N n^2 \chi_n^2 \rkk \,,
\end{equation}
where we define $\phi_n = \chi_n / a$ and introduce conformal time via ${\rm d}\eta = {\rm d}t/q$. 
The prime denotes differentiation with respect to $\eta$.
The action is then independent of the scale factor, and the corresponding classical equation of motion is 
\begin{equation}
 \chi_n'' + N^2 n^2 \chi_n = 0 \,. 
\end{equation}

We emphasize that, in the Lorentzian path integral formalism, the propagator itself does not directly represent the wave function. 
The transition amplitude from a vacuum state at $\eta = \eta_0$ with $\chi_n = \chi_{n,0}$ to $\eta = \eta_1$ with $\chi_n = \chi_{n,1}$ is given by 
\begin{equation}
 \Psi_n (\chi_{n,1}) = \int {\rm d} \chi_{n,0} G_{\chi_n} (\chi_{n,1}; \chi_{n,0}) \Psi_{0,\eta_0} (\chi_{n,0}) \,,
\end{equation}
where 
\begin{equation}
 G_{\chi_n} (\chi_{n,1}; \chi_{n,0}) = \int_{\chi_{n,0}}^{\chi_{n,1}} {\cal D}\chi_n e^{iS_n [\chi_n]} \,,
\end{equation}
and $\Psi_{0,\eta_0}(\chi_{n,0})$ denotes the initial wave function. 
We expect the initial wave function to contain only positive-frequency modes, as in the Minkowski vacuum (analogous to the Bunch--Davies vacuum state in inflationary cosmology), so that
\begin{equation}
 \Psi_{0,\eta_0} (\chi_{n,0}) = \sqrt{\frac{\gamma}{2\pi}} \exp \lkk - \frac{1}{2} \gamma_0 \chi_{n,0}^2 \rkk \,,
\end{equation}
with $\gamma_0 = n$. 
(See the Appendix for a detailed calculation in the general case and for a well-known example in a de Sitter background.) 
This choice is appropriate for the conformal scalar field, whose mode equation and corresponding wave function are independent of the scale factor, and thus should coincide with those in Minkowski spacetime.

Performing the Gaussian integral over $\chi_{n,0}$, we obtain the wave function at $\eta=\eta_1$ as 
\begin{equation}
 \Psi_n (\chi_{n,1}) \propto \sqrt{\frac{\gamma_1}{2\pi}} e^{- \frac{1}{2} \gamma_1 \chi_{n,1}^2}
 e^{-\frac{1}{2} \ln \lmk \frac{f'^*(\eta_1)}{f'^*(\eta_0)} \rmk } \,,
 \label{eq:Psib}
\end{equation}
where the mode function is
\begin{equation}
 f(\eta) = \frac{1}{\sqrt{2n}} e^{-i n N \eta} \,,
\end{equation}
and 
\begin{equation}
 {\rm Re} [\gamma_1] = n \,,
\end{equation}
(see \eq{eq:gammab1}). 
Thus, the wave function on any final time slice exhibits exponentially suppressed perturbations.

The second exponential factor in \eq{eq:Psib} arises from the functional determinant of fluctuations around the classical trajectory. 
Summing over all modes and including the degeneracy factor $n^2$, this contribution yields 
\begin{align}
 &\sum_n n^2 \lkk -\frac{1}{2} \ln \lmk \frac{f'^*(\eta_1)}{f'^*(\eta_0)} \rmk  \rkk
 \nonumber\\
 &= -\frac{i}{2} N (\eta_1 - \eta_0) \sum_n n^3 
 \nonumber\\
 &= i \int {\rm d}^4 x \sqrt{-g} \lmk \frac{1}{4 \pi^2 a^4} \sum_n n^3 \rmk  \,,
\end{align}
where we have used $N (\eta_1 - \eta_0) = \int N\, {\rm d}\eta$ and the volume factor $\int {\rm d}\Omega_3 = 2 \pi^2$. 
After regularizing the divergent sum, e.g., imposing a cutoff at the physical momentum $n_{\rm max}/a$, this term renormalizes the vacuum energy density $\Lambda$ ($=3 H^2$), as expected.

For a general scalar field (e.g., a massive or nonconformally coupled one), the same reasoning applies: 
the initial wave function should reduce to the positive-frequency form in the limit $q \to 0$, where differences due to mass or coupling become negligible. 
Hence, in such cases, the perturbations are also expected to be appropriately suppressed. 
We therefore conclude that scalar perturbations are generally suppressed in the Lorentzian path-integral formalism.

Finally, we note that the sign of the ``kinetic term" for $\phi$ is opposite to that for the scale factor, implying that the condition for a convergent Gaussian integral for fluctuations around the classical trajectory is reversed. 
As a result, the region of the lapse function yielding a convergent Gaussian integral is restricted to the real axis, $\Im N = 0$. 
The remaining question is how to define the conditionally convergent Lorentzian integral over the lapse function along the real axis. 
In the next section, we show that, owing to the additional regulator term, the integral becomes absolutely convergent for finite $\sigma$ and remains well defined in the limit $\sigma \to 0$.

%%%%%%%%%%%%%%%%%%%%%%%%%%%%%%%%%%%%%%%%%%%%%%%%%%%
\section{Path integral of the lapse function}
\label{sec:lapse}
%%%%%%%%%%%%%%%%%%%%%%%%%%%%%%%%%%%%%%%%%%%%%%%%%%%

We now consider the lapse integral 
\begin{equation}
 \tilde{\Psi}(q_1, \sigma) \equiv \int_0^\infty {\rm d}N \sqrt{\frac{3 \pi i}{2 (N + i 3 \pi^2 \sigma^2)}} e^{i S_g'[N; q_1,\sigma]}\,,
\end{equation}
with the action given by \eq{SgN}, which we hereafter denote as $S_g'[N; q_1, \sigma]$. 
We take the limit $\sigma \to 0$ after the integration.

\subsection{Behavior of the action on the real axis}

On the real axis $N\in\mathbb{R}$, $S_g'[N; q_1, \sigma]$ can be written explicitly as
\begin{equation}
\mathrm{Re}\big[i\,S_g'(N;q_1,\sigma)\big]
= -\,\frac{9\pi^4\sigma^2}{2}\;
\frac{\big(H^2 N - q_1/N\big)^{\!2}}
{\,1+\big(3\pi^2\sigma^2/N\big)^{\!2}}\,,
\quad (N\in\mathbb{R}) \,.
\label{eq:ReiS_real}
\end{equation}
In particular, for large $|N|$, we have 
\begin{equation}
\mathrm{Re}\!\left(i S_g'\right)
\;\sim\;
-\,\frac{9\pi^4}{2}\,H^4\,\sigma^2\,N^2 \,,
\qquad |N|\to\infty\,, \  N\in\mathbb{R} \,,
\label{eq:asymp}
\end{equation}
showing that the integrand is quadratically damped along the real axis.
More generally, along a horizontal line $N=x+i \, y$ with fixed $y$, one finds the large-$|x|$ behavior
\begin{equation}
\mathrm{Re}\!\left(i S_g'\right)
\;=\;
-\,\frac{3\pi^2 H^4}{2}\,\big(3\pi^2\sigma^2+y\big)\,x^2
\;+\;\mathcal{O}(|x|) \,,
\qquad |x|\to\infty \,,
\label{eq:asymp_line}
\end{equation}
which shows that the same quadratic damping persists whenever
\begin{equation}
\operatorname{Im}N \;=\; y \;>\; -\,3\pi^2\sigma^2 \,.
\label{eq:condition}
\end{equation}
Therefore, for any fixed $\sigma>0$, the lapse integral is well defined without invoking Picard--Lefschetz theory or deforming the contour into the upper half-plane.
All qualitative statements below remain valid when the limit $\sigma\to 0$ is taken \emph{after} the functional integrations, corresponding to the $q_0=0$ (“creation from nothing”) wave function explained in the Introduction.

\begin{figure}[t]
    \centering
    \includegraphics[width=1.0\linewidth]{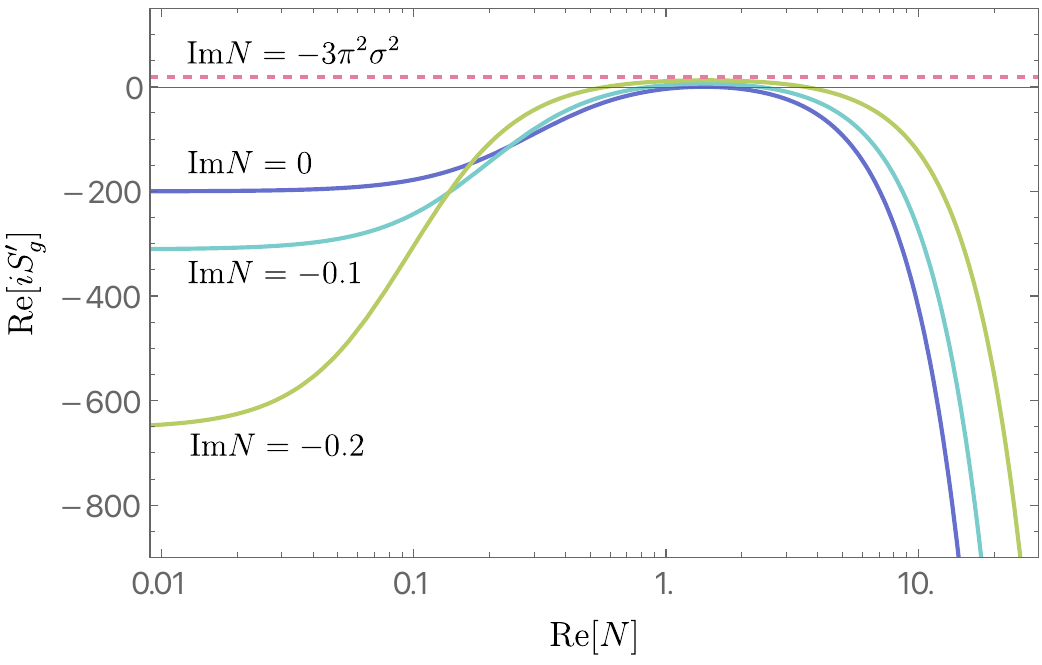}
    \caption{$\mathrm{Re}\!\left[i S_g'\right]$ versus $\mathrm{Re}\,N$ along $\operatorname{Im}N=0$ (blue solid), $-0.1$ (pale green solid), $-0.2$ (yellow solid), and $\operatorname{Im}N=-3\pi^2\sigma^2$ (red dashed), shown for $\sigma=0.1$. 
    The large-$|\mathrm{Re}\,N|$ suppression holds for all lines with $\operatorname{Im}N>-3\pi^2\sigma^2$, in agreement with \eqref{eq:asymp_line}.}
    \label{fig:1}
\end{figure}

Figure~\ref{fig:1} illustrates $\mathrm{Re}\!\left(i S_g'\right)$ as a function of $\mathrm{Re}\,N$ along $\operatorname{Im}N=0$ (blue), $-0.1$ (pale green), $-0.2$ (yellow), and $\operatorname{Im}N=-3\pi^2\sigma^2$ (red, dashed) for $\sigma=0.1$. 
As anticipated from \eqref{eq:asymp_line}, the integrand is strongly suppressed at large $|\mathrm{Re}\,N|$ whenever $\operatorname{Im}N>-3\pi^2\sigma^2$ ($\simeq-0.296$ for $\sigma=0.1$), while the leading quadratic damping vanishes precisely on the threshold.

\subsection{Numerical results of the path integral}

\begin{figure}[t]
    \centering
    \includegraphics[width=1.0\linewidth]{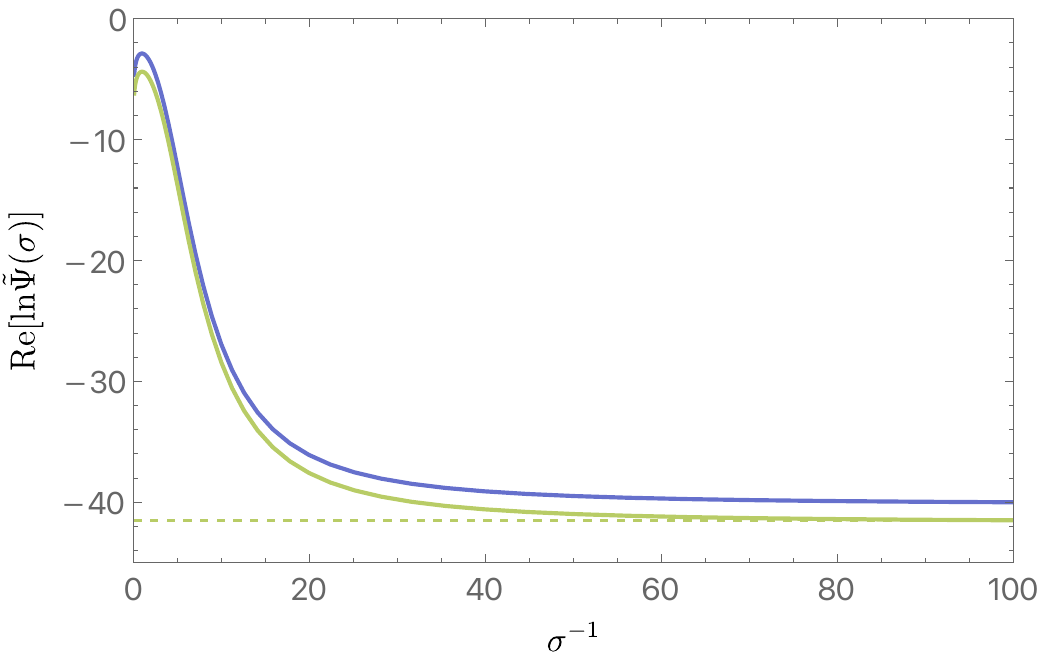}
    \caption{
    Results of the Lorentzian path integral as a function of $\sigma^{-1}$ for $H=1$ and $q_1 = 2$ (blue curve). The yellow curve represents the estimation obtained by the saddle-point approximation, and the dashed line indicates its asymptotic value in the limit $\sigma \to 0$.
    }
    \label{fig:1-2}
\end{figure}

We numerically evaluate the lapse function integral with a finite regulator $\sigma$.
The blue curve in Fig.~\ref{fig:1-2} shows the result of ${\rm Re}[\ln \tilde{\Psi}]$ as a function of $\sigma^{-1}$ for the case of $H=1$ and $q_1 = 2$.
We find that the result approaches a certain constant value in the limit $\sigma \to 0$.
This limiting value provides a definition of the conditionally convergent integral corresponding to the Lorentzian path integral in minisuperspace.

%-------------------------------------------------
\subsection{Saddle points}
%-------------------------------------------------

The path integrals over the scale factor and scalar field should be evaluated before performing the lapse integral, which initially requires $\Im N = 0$.
However, once the lapse integral is isolated, it can be evaluated by deforming the contour in the complex $N$–plane, allowing the use of the saddle-point approximation.

We deform the contour wherever analyticity and convergence permit (without crossing singularities).
The path integral can then be approximated by the saddle points lying on the deformed contour.
The saddle points $N_\star$ satisfy $\partial_N S_g'(N_\star)=0$, yielding four solutions:
\begin{equation}
N_\star^{(\pm,\pm)} \;=\; -\,i\,3\pi^2\sigma^2 
\;\pm\; \frac{\sqrt{H^2 q_1-1}}{H^2}
\;\pm\; i\,\frac{\sqrt{\,1+9\pi^4 H^4\sigma^4\,}}{H^2} \,,
\label{eq:saddles}
\end{equation}
valid for $q_1>H^{-2}$.
In the limit $\sigma\to 0$, these approach
$N_\star^{(\pm,\pm)} \to \pm H^{-2}\!\left(\sqrt{H^2q_1-1}\,\pm i\right)$.

\begin{figure}[t]
    \centering
    \includegraphics[width=1.0\linewidth]{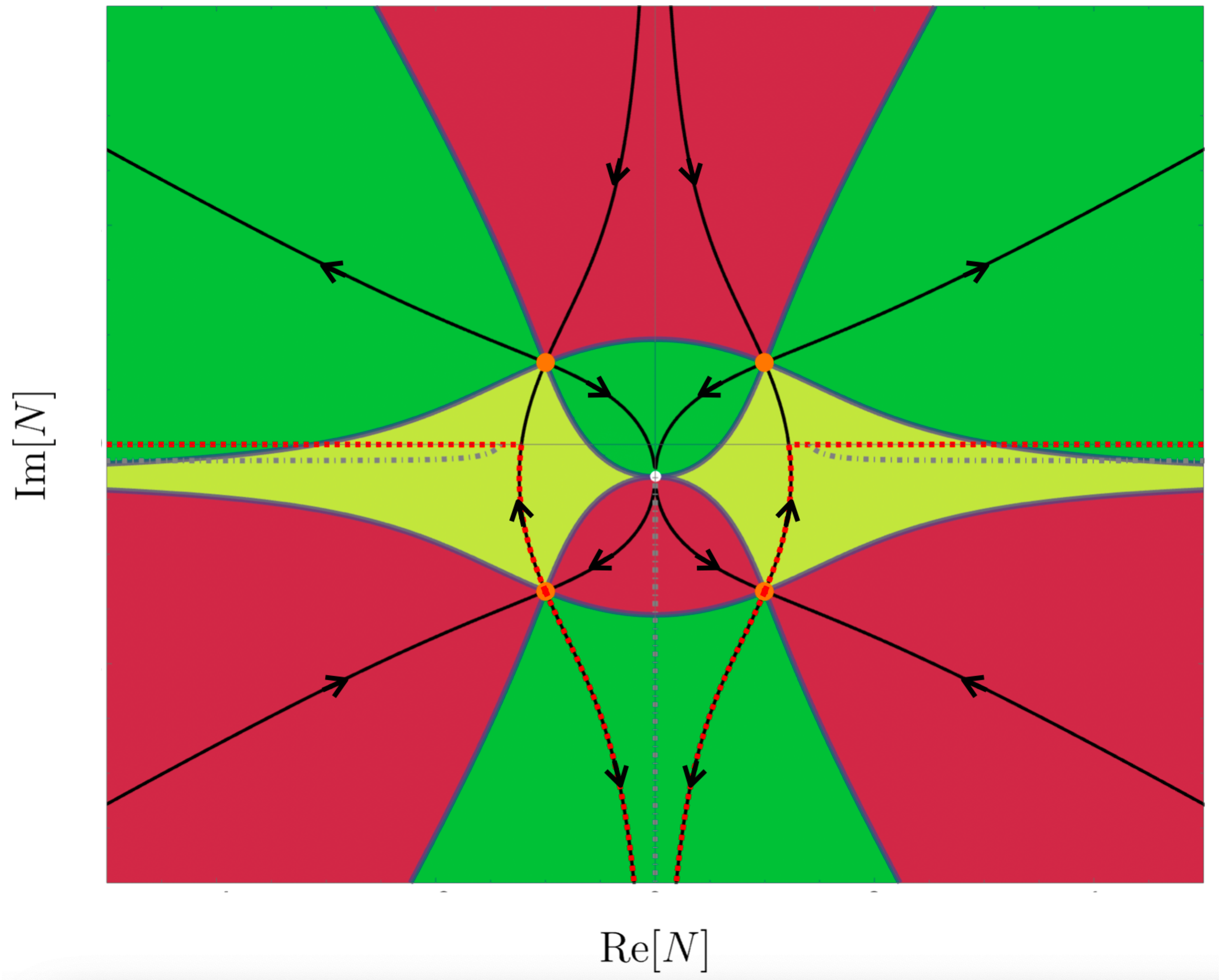}
    \caption{Convergence diagram in the complex $N$-plane. 
    Green (red) shading indicates asymptotic regions of convergence (divergence). 
    Orange dots denote the four saddle points $N_\star^{(\pm,\pm)}$ and the white dot marks the singularity at $N=-\,i\,3\pi^2\sigma^2$. 
    Black curves are steepest–descent/ascent lines (arrows indicate descent). 
    The HH wave function is obtained by considering an integration contour from $-\infty$ to $+\infty$ that passes below the singularity (red dotted path), which remains entirely in $\operatorname{Im}N\le 0$ and is absolutely convergent. The gray dot-dashed curves represent contours along which the scale factor temporarily vanishes at a finite time.}
    \label{fig2}
\end{figure}

Figure~\ref{fig2} displays the four saddles (orange dots) together with the singularity at $N=-\,i\,3\pi^2\sigma^2$ (white dot). 
Steepest–descent/ascent curves are shown in black, with arrows indicating the direction of descent. 
Depending on the choice of the original contour, a subset of these saddles contributes.
For the Lorentzian path integral, where the original contour is $N \in (0,\infty)$ along the real axis, the saddle point $N_\star^{(+,+)}$ contributes dominantly~\cite{Feldbrugge:2017kzv,Feldbrugge:2017fcc}.

The yellow curve in Fig.~\ref{fig:1-2} shows the result obtained by the saddle-point approximation, which agrees with our numerical results up to an $\mathcal{O}(1)$ numerical factor.
The dashed line indicates the asymptotic value of the saddle-point result in the limit $\sigma \to 0$.
In this limit, we approximately obtain the tunneling wave function
\begin{equation}
 \Psi(q_1) \propto \exp \lkk - \frac{4\pi^2}{H^2} \lmk 1  - i \lmk H^2 q_1 - 1 \rmk^{3/2} \rmk \rkk \,,
\end{equation}
emerging from the purely Lorentzian path integral.
We emphasize that, although our results are quantitatively consistent with those obtained via the Picard--Lefschetz prescription~\cite{Feldbrugge:2017kzv,Feldbrugge:2017fcc}, the conceptual basis of our approach is distinct.

%%%%%%%%%%%%%%%%%%%%%%%%%%%%%%%%%%%%%%%%%%%%%%%%%%%
\section{Modified contour for the HH wave function}
\label{sec:HH}
%%%%%%%%%%%%%%%%%%%%%%%%%%%%%%%%%%%%%%%%%%%%%%%%%%%

We next consider reproducing the HH wave function by considering a different contour of original integration.
First, we disregard the conformal factor issue and assume the semiclassical approximation for the path integral, particularly for the scale factor.
In other words, we no longer impose the condition $\Im N \ge 0$.
However, we do take the path integral over scalar perturbations seriously, since quantum field theory is well established even in a curved spacetime background.
Specifically, the Gaussian integral for fluctuations converges only in the region $\operatorname{Im}N \le 0$, and we continue to respect this condition throughout this section.
Second, we consider a ``nearly Lorentzian" contour for the lapse function, such that the integration path in the complex $N$-plane approaches the real axis asymptotically as $\abs{N} \to \infty$.

Here we note that 
a common approach in the literature is to evaluate the path integral over scalar perturbations \emph{after} those over the lapse function and the scale factor, assuming negligible backreaction throughout the entire complex $N$ plane (see, e.g., Ref.~\cite{Feldbrugge:2017fcc}).
However, it is important to note that the Gaussian functional integral for fluctuations of $\phi_n$ converges only when ${\rm Im} N \le 0$. Otherwise, the quadratic term flips sign and the integral becomes an inverse Gaussian. 
Accordingly, the lapse integrand that includes perturbations is analytic only in the domain $\operatorname{Im}N \le 0$.
In other words, when deforming the contour in the complex $N$–plane, we must restrict ourselves to the region $\operatorname{Im}N \le 0$.
This restriction plays a central role in the contour choices used to evaluate the full path integral and to construct the HH wave function.

\subsection{Saddle points}

To obtain the HH wave function, we choose an original contour running from $-\infty$ to $+\infty$ that passes \emph{below} the singularity near the origin (located at $N\simeq -\,i\,3\pi^2\sigma^2$ for $\sigma>0$).%
\footnote{
See Ref.~\cite{Banihashemi:2024aal} for an equivalent choice of contour discussed in a different context.
}
The steepest-descent deformation in the lower half-plane then selects the two saddles with negative imaginary parts, $N_\star^{(\pm,-)}$ (see the red dotted curves in Fig.~\ref{fig2}).
In the limit $\sigma\to 0$,
their combined contribution yields the HH wavefunction 
\begin{equation}
\Psi_{\rm HH}(q_1)
\;\propto\;
\exp\!\left(+\,\frac{4\pi^2}{H^2}\right)\,
\cos\!\left[\, \frac{4\pi^2}{H^2} \left( H^2 q_1-1\right)^{3/2}\right] \,.
\label{eq:HH}
\end{equation}
The entire contour deformation stays within $\operatorname{Im}N \le 0$, ensuring that the path integral of scalar and tensor perturbations remains convergent.
This prescription also avoids the problem of unsuppressed fluctuations that arises when a merely conditionally convergent integral must be deformed into $\operatorname{Im}N > 0$ in the absence of an initial wave function for the scalar/tensor perturbations.

A practical advantage of the HH contour is that the contributing saddles lie entirely in the lower half-plane.
Consequently, the scalar/tensor perturbation integrals remain Gaussian and can be evaluated at the saddle point of $N$ after the $N$-integration, even when one adopts the Dirichlet boundary condition (i.e., $\phi = 0$ at the initial boundary) for these perturbations without the initial wave function.
In this setup, the positive-frequency modes are consistent with the vanishing initial condition for the perturbations, yielding properly suppressed fluctuations (see, e.g., Ref.~\cite{DiazDorronsoro:2017hti}).
Thus, one can safely perform the path integral for general scalar and tensor perturbations after integrating over the lapse function and the scale factor, provided the backreaction remains negligible.

%-------------------------------------------------
\subsection{Singular curve}
%-------------------------------------------------

For certain complex values of $N$, the classical history $q_c(t)$ can pass through $q_c(t)=0$ at an intermediate time $t\in(0,1)$.
At such points, higher-curvature corrections may become significant, and the validity of the on-shell action may be questionable~\cite{DiTucci:2019dji}.
These points can be identified in the complex $N$–plane as the solutions of
\begin{equation}
q_c(t)=0\qquad (0<t<1) \,.
\end{equation}
Using \eq{qt} and multiplying by $(N+i\,3\pi^2\sigma^2)$ to eliminate the denominator, this condition reduces to the cubic equation
\begin{align}
H^2(t^2-t)\,N^3
\;+\; i\,3\pi^2\sigma^2 H^2(t^2-1)\,N^2
\nonumber \\
\;+\; q_1\,t\,N
\;+\; i\,3\pi^2\sigma^2 q_1
\;=\;0 \,.
\label{eq:branch_cubic}
\end{align}
For each fixed $t\in(0,1)$ the roots of \eqref{eq:branch_cubic} define three curves $N = N_p^{(i)}(t)$ ($i=1,2,3$) in the complex $N$-plane.

Setting $N=-iY$ with $Y>0$ transforms \eqref{eq:branch_cubic} into a cubic equation for the real variable $Y$, which ensures the existence of a purely imaginary root $N_p^{(1)}(t)=-iY(t)$. 
The remaining two curves, $N_p^{(2)}(t)$ and $N_p^{(3)}(t)$, form an off-axis conjugate pair with equal imaginary parts and opposite real parts.

The asymptotic behavior of these curves is obtained as
\begin{align}
&N_p^{(1)}(t \to 0^+) \approx -\,i\,3\pi^2\sigma^2 \frac{1}{t} \,,
\\
&N_p^{(2,3)}(0)=\pm \sqrt{q_1/H^2} \,, 
\\
&N_p^{(1)}(1)=-\,i\,3\pi^2\sigma^2  \,,
\\
&N_p^{(2,3)}(t \to 1^{-}) \approx 
 \pm \frac{q_1}{\sqrt{H(1-t)}} - \frac{i 3\pi^2\sigma^2}{2} \,.
\end{align}
The singular curve $N_p^{(1)}(t)$ thus runs down the negative imaginary axis from $-i\infty$ (as $t\to0^+$) to $-i\,3\pi^2\sigma^2$ (as $t\to1^{-}$). The other two singular curves extend to complex infinity in the lower half-plane as $t\to1^{-}$.

\begin{figure}[t]
    \centering
    \includegraphics[width=1.0\linewidth]{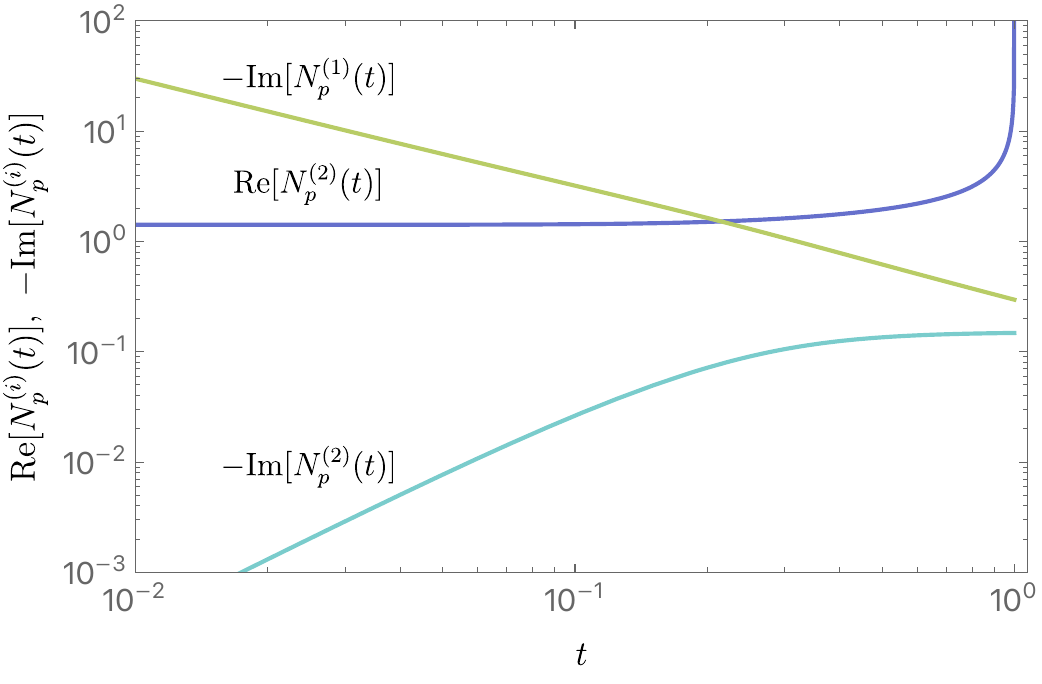}
    \caption{
    Real and imaginary parts of the solutions $N_p^{(i)}(t)$ of \eqref{eq:branch_cubic} for $H=1$, $q_1=2$, and $\sigma=0.1$. 
    }
    \label{fig3}
\end{figure}

Figure~\ref{fig3} shows $\mathrm{Re}\,N_p^{(2)}(t)$ (blue), $-\mathrm{Im}\,N_p^{(2)}(t)$ (pale green), and $-\mathrm{Im}\,N_p^{(1)}(t)$ (yellow) for $H=1$, $q_1=2$, and $\sigma=0.1$. 
Note that $\mathrm{Re}\,N_p^{(1)}(t)=0$ for all $t$, while $N_p^{(3)}(t)$ satisfies $\mathrm{Re}\,N_p^{(3)}(t)=-\mathrm{Re}\,N_p^{(2)}(t)$ and $\mathrm{Im}\,N_p^{(3)}(t)=\mathrm{Im}\,N_p^{(2)}(t)$. 
These results are consistent with the analytic discussion above.

The contours of $N_p^{(i)}(t)$ are shown in Fig.~\ref{fig2} as gray dot-dashed curves.
The HH contour (running from $-\infty$ to $+\infty$ and passing below the singularity at $N=-\,i\,3\pi^2\sigma^2$) can be chosen to avoid the two off-axis singular curves. 
The on-axis singular curve $N_p^{(1)}(t)$ cannot be avoided because it runs from $-i \infty$ to the singularity. 
However, large negative $\operatorname{Im}N$ corresponds to the limit of $t \to 0^+$ and becomes indistinguishable from the $q_c(0)=0$ initial condition, particularly for a small $\sigma$. 
Indeed, in the limit $\sigma=0$, this family collapses to the singular point at $N = - 3 \pi^2 \sigma^2 i = 0$ and effectively disappears. 
Consequently, we expect that the HH contour may cross the imaginary axis far below the singularity without encountering any physical obstruction. 

%%%%%%%%%%%%%%%%%%%%%%%%%%%%%%%%%%%%%%%%%%%%%%%%%%%
\section{Discussion and conclusion}
\label{sec:discussions}
%%%%%%%%%%%%%%%%%%%%%%%%%%%%%%%%%%%%%%%%%%%%%%%%%%%

We have proposed a regulator for the Lorentzian minisuperspace path integral that renders the lapse integral convergent along (or slightly below) the real axis.
The tunneling wave function is successfully reproduced from a purely Lorentzian path integral, even beyond the semiclassical approximation for both the scale factor and a conformally coupled massless scalar field.
When scalar and tensor perturbations are included, one should specify their initial wave function in order to compute the wave function at a final time slice using the Lorentzian propagator.
This initial state should be compatible with the Bunch–Davies vacuum so that each mode function has positive frequency, ensuring that the perturbations remain suppressed.

If one instead adopts a modified contour for the lapse function and assumes the semiclassical approximation for the scale factor, the HH wave function can be reproduced.
Because the integral remains within the region $\Im N \le 0$, the linear perturbations are still Gaussian and suppressed, 
even when they are evaluated after performing the path integrals over the lapse function and the scale factor, without introducing an explicit initial wave function.

Our regulator admits a simple physical interpretation:
before taking the limit $\sigma\to 0$, the Gaussian factor effectively prepares the universe in a narrow wave packet for the initial size $q_0$, corresponding to a small quantum uncertainty in the scale factor.
In this sense, the universe originates from a quantum fluctuation rather than from absolute “nothing,” provided that $\sigma$ remains finite.
Equivalently, the variational principle leads to the Robin boundary condition \eqref{eq:Robin} at $t=0$, so the regulator can also be interpreted as a boundary-condition prescription for the wave function of the universe.

\section*{Acknowledgments}
This work was supported by JSPS KAKENHI Grant Number 23K13092.

\section*{Appendix}

In this Appendix, we reexamine the calculation of scalar perturbations with a time-dependent frequency using the propagator in the Lorentzian path integral formalism.
We then apply the result to the case of scalar perturbations in a de Sitter background.
Although this is not directly related to the minisuperspace model considered in the main text, it illustrates the importance of the initial wave function for perturbations and clarifies how the standard results in the literature can be recovered within this approach.

\subsection{General formalism for a harmonic oscillator}

We first consider, in general, a quantum mechanical system of a harmonic oscillator with a time-dependent frequency:
\begin{equation}
 \mathcal{L}_k (\chi_k, \chi_k') = \frac{1}{2} \lmk \chi_k'^2 - \omega_k^2 (\eta) \chi_k^2 \rmk \,,
\end{equation}
where $\eta$ denotes the time variable, and the prime represents differentiation with respect to $\eta$.
The meaning of the parameters in the de Sitter background will be specified shortly, but for now we treat $\omega_k(\eta)$ as a general function.
We decompose $\chi_k$ into a classical part $\chi_{\rm cl}$ and a fluctuation $\delta \chi$ as
\begin{equation}
 \chi_k (\eta) = \chi_{\rm cl} (\eta) + \delta \chi (\eta) \,.
 \label{eq:decompose}
\end{equation}
The classical part satisfies the equation of motion such as 
\begin{equation}
 \chi_{\rm cl}'' + \omega_k^2 \chi_{\rm cl} = 0 \,.
\end{equation}
We denote two independent solutions of this equation by $u$ and $v$, satisfying the boundary conditions 
\begin{align} 
 u(\eta_0) = 0, \qquad u'(\eta_0) = 1 \,,
 \label{eq:bc1}
 \\
 v(\eta_0) = 1, \qquad v'(\eta_0) = 0\,,
 \label{eq:bc2}
\end{align}
at an initial time $\eta = \eta_0$. 
The general solution with Dirichlet boundary conditions $\chi_{\rm cl} (\eta_0) = \chi_{k,0}$ at $\eta = \eta_0$ and $\chi_k (\eta_1) = \chi_{k,1}$ at $\eta = \eta_1$ is then written as 
\begin{equation}
 \chi_k (\eta) = \chi_{k,0} v(\eta) + \frac{\chi_{k,1} - \chi_{k,0} v (\eta_1)}{u (\eta_1)} u(\eta) \,.
\end{equation}

We now introduce a mode function $f(\eta)$ that satisfies the equation of motion and is normalized by the Wronskian condition
\begin{equation}
 W \equiv  f'^* (\eta) f (\eta)  - f' (\eta)  f^* (\eta)  = i \,.
\end{equation}
The functions $u$ and $v$ can be expressed in terms of $f(\eta)$ as 
\begin{align}
 &u(\eta) = \frac{1}{i} \lmk f^* (\eta) f(\eta_0) - f(\eta) f^* (\eta_0) \rmk \,,
 \\
 &v(\eta) = \frac{1}{i} \lmk f(\eta) f'^* (\eta_0) - f^* (\eta) f'(\eta_0) \rmk \,.
\end{align}

With the decomposition of \eqref{eq:decompose}, the path integral from $\chi_k(\eta_0) = \chi_{k,0}$ to $\chi_k (\eta_1) = \chi_{k,1}$ is given by 
\begin{align}
 &G_{\chi_k} (\chi_{k,1},\eta_1; \chi_{k,0}, \eta_0) \nonumber\\
 &= 
 \int_{\chi_k(\eta_0) = \chi_{k,0}}^{\chi_k (\eta_1) = \chi_{k,1}} 
 \mathcal{D} \chi_k \exp \lkk i \int_{\eta_0}^{\eta_1} {\rm d} \eta \mathcal{L}_k \rkk 
 \nonumber\\
 &= e^{i S_{\rm cl}} 
 \int_{\delta \chi (\eta_0) = 0}^{\delta \chi (\eta_1) = 0} 
 \mathcal{D} \delta \chi \exp \lkk i \int_{\eta_0}^{\eta_1} {\rm d} \eta \mathcal{L}_k (\delta \chi, \delta \chi') \rkk 
 \nonumber\\
 &\propto \lkk {\rm det} \lmk - \frac{{\rm d}^2}{{\rm d} \eta^2 } - \omega_k^2  \rmk \rkk^{-1/2} e^{i S_{\rm cl}} \,.
\end{align}
The classical action is given by 
\begin{equation}
 S_{\rm cl} = \frac{1}{2} 
 \lkk \frac{u'(\eta_1)}{u(\eta_1)} \chi_{k,1}^2 + 
 \frac{v(\eta_1)}{u(\eta_1)} \chi_{k,0}^2 
 - \frac{2}{u(\eta_1)} \chi_{k,0} \chi_{k,1}
 \rkk \,.
\end{equation}
The determinant in the prefactor can be evaluated using the Gel'fand–Yaglom theorem as~\cite{Dunne:2007rt}
\begin{equation}
{\rm det} \lmk - \frac{{\rm d}^2}{{\rm d} \eta^2 } - \omega_k^2  \rmk = u (\eta_1) \,.
\end{equation}
These results provide the full expression for the propagator of a harmonic oscillator with a general time-dependent frequency.

One may take the initial quantum state to be the vacuum at a given time $\eta = \eta_0$. For this purpose, we represent the quantum operator $\hat{\chi}_k$ in terms of the annihilation and creation operators as
\begin{equation}
 \hat{\chi}_k = \hat{a} f (\eta) + \hat{a}^\dagger f^* (\eta) \,,
\end{equation}
where $[\hat{a}, \hat{a}^\dagger] = 1$. 
Noting that the conjugate momentum is given by $\hat{\pi}_k = \hat{q}_k'$, 
we obtain 
\begin{equation}
 \hat{a}(\eta) = i \lmk f^* (\eta) \hat{\pi}_k - f'^* (\eta) \hat{\chi}_k \rmk \,.
\end{equation}
Defining the vacuum state by $\hat{a} \left\vert 0 \right>_{\eta_0} = 0$ and replacing $\hat{\chi}_k \to \chi_k$ and $\hat{\pi}_k \to - i \del_{\chi_k}$, the vacuum wave function $\Psi_{0,\eta_0}$ satisfies 
\begin{equation}
 \lmk f^* \del_{\chi_k}  - i f'^* \chi_k \rmk \Psi_{0,\eta_0} = 0 \,.
\end{equation}
This yields
\begin{equation}
 \Psi_{0,\eta_0} (\chi_{k,0}) = \sqrt{\frac{\gamma_0}{2\pi}} \exp \lkk - \frac{1}{2} \gamma_0 \chi_{k,0}^2 \rkk \,,
\end{equation}
where 
\begin{equation}
 \gamma_0 \equiv - i \frac{f'^* (\eta_0)}{f^* (\eta_0) } \,.
 \label{eq:gamma}
\end{equation}

The transition amplitude from the vacuum state at $\eta = \eta_0$ to a later time $\eta = \eta_1$ is then given by
\begin{align}
 \Psi (\chi_{k,1}) = \int {\rm d} \chi_{k,0} G_{\chi_k} (\chi_{k,1},\eta_1; \chi_{k,0}, \eta_0) \Psi_{0,\eta_0} (\chi_{k,0}) 
 \\
 \propto \sqrt{\frac{\gamma_0}{u(\eta_1)}} 
 \int {\rm d} \chi_{k,0} \exp \lkk i S_{\rm cl} - \frac{1}{2} \gamma_0 \chi_{k,0}^2 \rkk \,.
\end{align}
We can rewrite 
\begin{equation}
 i S_{\rm cl} - \frac{1}{2} \gamma_0 \chi_{k,0}^2 
 = \frac{i}{2} \frac{f^* (\eta_1)}{f^* (\eta_0) u(\eta_1)} \lmk \chi_{k,0} -  \frac{f^* (\eta_0)}{f^* (\eta_1)} \chi_{k,1} \rmk^2 
 - \frac{1}{2} \gamma_1 q_1^2 \,,
\end{equation}
with
\begin{equation}
 \gamma_1 = - i \frac{f'^* (\eta_1)}{f^* (\eta_1) }\,,
 \label{eq:gammab2}
\end{equation}
where we have used the relation $v(\eta_1) i - \gamma_0 u(\eta_1) = i f^* (\eta_1) / f^* (\eta_0)$. Since the exponent is quadratic in $\chi_{k,0}$, the Gaussian integral can be evaluated analytically, yielding
\begin{equation}
 \Psi (\chi_{k,1}) \propto 
 \sqrt{\frac{\gamma_1}{2\pi}} e^{- \frac{1}{2} \gamma_1 \chi_{k,1}^2} e^{-\frac{1}{2} \ln \lmk \frac{f'^*(\eta_1)}{f'^*(\eta_0)} \rmk }\,.
 \label{eq:gammab1}
\end{equation}
The second exponential factor represents the functional determinant for quantum fluctuations, while $\gamma_1$ characterizes the inverse width of the fluctuation at $\eta = \eta_1$. The final wave function is Gaussian and remains suppressed when the mode function retains positive-frequency.

\subsection{Perturbations in de Sitter background}

As a well-known application,
we apply the above formalism to scalar perturbations in a de Sitter background.
We employ the conformal time coordinate and metric
\begin{equation}
{\rm d}s^2 = a^2(\eta) \lmk - {\rm d} \eta^2 + {\rm d} {\bm x}^2 \rmk \,,
\end{equation}
with $a(\eta) = - 1/(H \eta)$.
In the previous formulation, $\chi_k$ is identified with the Fourier modes of the original field via $\chi_k (\eta) = a(\eta) \phi_k (\eta)$,
and
\begin{equation}
 \omega_k^2 (\eta) = k^2 + \frac{1}{\eta^2} \lmk \frac{m^2}{H^2} + 12 \xi - 2 \rmk \,.
\end{equation}
We define $\nu^2 - 1/4 \equiv 2 - m^2 / H^2 - 12 \xi$,
where $\xi$ denotes the non-minimal coupling constant.
In this case, the solution to the equation of motion can be expressed in terms of the Hankel functions $H_\nu^{(i)}$ ($i=1,2$).
Since $f(\eta)$ represents the mode function with positive frequency in the subhorizon limit, namely $f(\eta) = e^{-i k \eta}/\sqrt{2k}$ for $k \eta \to - \infty$, as in Minkowski spacetime, 
it is natural to take
\begin{align}
 f(\eta) = \frac{\sqrt{\pi \eta}}{2} H_\nu^{(1)} ( - k \eta) \,,
 \\
 f^*(\eta) = \frac{\sqrt{\pi \eta}}{2} H_\nu^{(2)} ( - k \eta) \,,
\end{align}
as the mode functions. 
This choice corresponds to the Bunch-Davies vacuum. We then obtain $\gamma_0 = k$ from \eq{eq:gamma}, reproducing the inverse width of quantum fluctuations in Minkowski spacetime in the subhorizon limit. Note that the mode function of the original field, $\phi_k = \chi_k / a$, diverges in the limit $\eta \to -\infty$, whereas the wave function itself remains finite.% 
\footnote{ Strictly speaking, the second exponential factor in \eq{eq:gammab1} contains a UV divergence that must be properly regulated. Here we simply emphasize that the wave function from the classical part remains finite once the initial wave function is included. }

Substituting these expressions into \eq{eq:gammab2}, we obtain 
\begin{equation}
 {\rm Re} [\gamma_1] \approx \frac{2^{1-\nu} \pi}{\lmk \Gamma (\nu) \rmk^2} k \lmk - k \eta \rmk^{2\nu - 1} \,,
\end{equation}
for $\nu >0$ 
in the superhorizon limit of $\abs{k \eta} \to 0$, 
where $\Gamma(\nu)$ is the Gamma function. 
The numerical prefactor can be approximated as $1 - 2 \lkk \ln 2 + \Gamma' (3/2)/\Gamma(3/2) \rkk \lmk \nu - \frac{3}{2} \rmk$ for $\abs{\nu - 3/2} \ll 1$. 
For $\nu = 3/2$, 
we obtain ${\rm Re} [\gamma_1] = k^3 \eta^2$, implying that the variance of the fluctuation of $\phi_k$ ($= \chi_k / a$) is $H^2 / k^3$. This reproduces the standard scale-invariant spectrum of quantum fluctuations during slow-roll inflation.

The lessons from this well-known example are as follows: (i) it is not problematic that $\phi_k$ diverges in the limit $\eta \to -\infty$; (ii) the initial wave function must be incorporated in order to compute the wave function at a later time using the propagator; and (iii) it is natural to adopt the positive-frequency mode function in the limit $\eta \to -\infty$, corresponding to the Bunch-Davies vacuum. We expect these features to persist for perturbations in the Lorentzian path integral formalism applied to the minisuperspace model in quantum cosmology. In particular, these conditions should hold for a massless conformally coupled scalar field, which is insensitive to the scale factor even as $q \equiv a^2 \to 0$. Since the equation of motion for a general scalar field reduces to that of a massless conformal field in the limit $q \to 0$, it is reasonable to assume that the same initial wave function applies also to general scalar fields.

Finally, we comment on the Euclidean path integral formulation. In this formalism, the initial wave function need not be introduced explicitly to reproduce the Bunch-Davies vacuum. The mode function is chosen to be regular in the limit $\abs{\eta} \to \infty$, ensuring that the action remains regular (see, e.g., Ref.~\cite{Maldacena:2024uhs}). The corresponding solution is unique in the de Sitter background. Thus, the wave function at a final boundary surface can be computed solely from the propagator without specifying the initial wave function, yielding a result consistent with \eq{eq:gammab1}.

The Euclidean path integral formulation can be generalized to the case of a complex lapse function with ${\rm Im} N < 0$. However, for ${\rm Im} N \ge 0$, the formalism fails to reproduce the suppressed fluctuations characteristic of the de Sitter background.% 
\footnote{ If one adopts the $-i\epsilon$ prescription, the formalism can be extended even to the case ${\rm Im} N = 0$. However, note that this prescription corresponds to a small negative value of ${\rm Im} N$. } 
In particular, even for a massless conformally coupled scalar field, the resulting fluctuation is inconsistent with that obtained after a conformal transformation to Minkowski spacetime. This implies that such an Euclidean-like formalism cannot be justified in the region ${\rm Im} N \ge 0$. We therefore insist that, in the case of ${\rm Im} N \ge 0$, the initial wave function must be included, as in the purely Lorentzian path integral formalism described above. 
Including the initial wave function of the form \eq{eq:gamma}, the calculation of the total wave function becomes identical to that discussed in Ref.~\cite{Vilenkin:2018dch,Vilenkin:2018oja}. The initial wave function leads to a Robin boundary condition for the mode functions of the perturbations, which selects the appropriate mode ensuring that the resulting wave function represents suppressed perturbations. We thus conclude that the perturbations remain suppressed in the region ${\rm Im} N \ge 0$, provided that the initial wave function is appropriately included.

\bibliography{reference}

\end{document}